\documentclass[a4paper,fleqn]{cas-sc}
\usepackage[numbers]{natbib}

\def\tsc#1{\csdef{#1}{\textsc{\lowercase{#1}}\xspace}}
\tsc{MGSE}

\def \b {\begin{eqnarray}}
\def \el#1 {\label{#1} \e}
\def \e {\end{eqnarray}}
\def \ds {\displaystyle }
\def \no {\nonumber\\ }
\newcommand{\ex}[1]{\,\mathrm{e}^ {\ds{#1}}}
\newcommand{\ave}[1]{\langle {#1} \rangle}
\renewcommand{\vec}[1]{\mathbf{#1}}
\newcommand{\abs}[1]{\left| {#1} \right|}

\begin{document}
\let\WriteBookmarks\relax
\def\floatpagepagefraction{1}
\def\textpagefraction{.001}
\shorttitle{Velocity autocorrelations spectra in G/W mixtures}
 
\title[mode = title]{Molecular velocity auto-correlations in glycerol/water mixtures studied  by  NMR MGSE method}
\tnotemark[1]
\tnotetext[1]{This document is the results of the research
   program funded by  the Slovenian research agency, ARRS, under the program P1-0060, \lq\lq{}Experimental biophysics of complex systems and imaging in bio-medicine\rq\rq{}.}
 \author[1]{Janez Stepi\v{s}nik}[ orcid=0000-0002-2930-5604]
 \cormark[1]
\fnmark[1]
\credit{Experimental biophysics of complex systems and imaging in bio-medicine}
\address[1]{University of Ljubljana, Faculty of Mathematics and Physics, Physics Department, Jadranska 19, 1000 Ljubljana, Slovenia}                         
\author[3]{Carlos Mattea}
\author[3]{Siegfried Stapf} 
\address[3]{Dept.Technical Physics II, TU Ilmenau, 98684 Ilmenau, Germany, EU}
\author[1,2]{Ale\v{s} Mohori\v{c}}
 \address[2]{Institute Jo\v{z}ef Stefan, Jamova 39, 1000 Ljubljana, Slovenia, EU}

\cortext[cor1]{Corresponding author}
\nonumnote{All authors participated in the measurements, the first author and partly the last one participated in the analysis and interpretation of the data and drafting the manuscript. We acknowledge also the contribution of prof. I. Ser\v{s}a and dr. F. Bajd from Josef Stefan Institute to assist in the preparation of experiments on the $100$ MHz NMR device. }

\begin {abstract}
Molecular dynamics in  binary mixtures of water and glycerol  was studied by measuring the spectrum of water velocity auto-correlation in  the frequency range from  $0.05-10$ kHz  by using  the NMR method of modulated gradient spin echo.  The method shows that the  diversity of diffusion signature in the short spin trajectories provides  information about heterogeneity of molecular motion due to the motion in the micro-vortexes of hydrodynamic  fluctuation, which is especially pronounced for the mixtures with  low glycerol content.  As concentration of glycerol increases above $10$vol$\%$, a new feature of spectrum appears  due  to interaction of water  molecules  with the clusters formed around hydrophilic  glycerol molecules. New spectrum  exposes a rate  thickening of molecular friction, according to   Einstein-Smoluchowski-Kubo formula, which inhibits rapid molecular motions and  creates the conditions  for a slow process of spontaneously  folding  of disordered poly-peptides into  biologically active protein molecules when  immersed in such a  mixture.
\end{abstract}



\begin{keywords}
molecular dynamics \sep glycerol/water mixture  \sep NMR \sep gradient spin-echo \sep self-diffusion \sep molecular velocity auto-correlation \sep shear rate viscosity thickening  \sep protein folding 
\end{keywords}
 
\maketitle

\section{Introduction}
Earliest models   of liquids as totally disordered structures have been replaced  by  models of systems with a long range-disordered and short range-ordered systems where  molecules can associate in clustes~\cite{Gadre,Guo} due to  intermolecular interactions~\cite{Vaitheeswaran,Saykally}, but there  is still lack of understanding  of molecular dynamic  that plays an important role in biological systems. It  refers primarily to liquids with  hydrogen bonding such as water, alcohol, glycerol and  mixtures thereof. The molecular mechanisms of these liquids, which spontaneously folds a disordered poly-peptide   into the unique structure  of a biologically active protein molecule, when imersed in them,  are still  not   understood.  Particularly pronounced in maintaining the structure of biologically active macromolecules and promoting protein self-assembly are glycerol-water (G/W) mixtures~\cite{Dashnau}. Despite considerable research efforts~\cite{Hansen,Chaikin} these questions remain among the key unresolved issues in soft condensed matter physics, physical chemistry, materials science and biophysics.

 The G/W mixtures have already been a subject of extensive research involving thermodynamic measurements~\cite{Marcus}, broadband dielectric measurement~\cite{Hayashi,Puzenko}, NMR~\cite{Doess,Mizuno,DErrico},  infrared (IR)~\cite{Dashnau}, and Raman  spectroscopy~\cite{Mendelovici}, etc. These studies reveal changes in the aqueous structure beyond the first neighbor level, but general properties of the G/W mixtures and how they  affect the molecular dynamics are not revealed. They also show that macrophages are formed, because  the glycerol is  a small molecule of trihydric alcohol with  large affinity to form 3D hydrogen-bonded network pervading the bulk of  mixtures~\cite{Chen,Popov,Murata}. Hydrogen bond energy between glycerol molecules  ($0.25$ eV) and between water ($0.19$ eV )   are lower than the binding energy between water and glycerol molecules ($0.29$ eV), leading to the molecular clusterization~\cite{Efimov}  confirmed by the dielectric  measurements with  almost tri-fold increase of the activation enthalpy of $0.9$ mol$\%$ G/W mixture compared to the   pure water at $H=16.2$ kJ/mol ~\cite{Behrends}. 
 
In liquids the thermal molecular motion  is impeded by  interactions with its neighbors.  Velocity auto-correlation function (VAF) is a quantity containing information about the underlying processes of molecular interaction  and dynamics~\cite{Kubo2}.  The VAF, which is associated with a number of physical properties, such as thermal and mass diffusion, sound propagation, transverse-wave excitation, and can have either a single-particle or a collective nature, has  a profound meaning in the statistical physics of fluids~\cite{Hansen, Chaikin, Pozar}, but  it is extremely  difficult to measure. Some information can be obtained by  neutron scattering~\cite{Sakamoto,Larsson,Ardente} and light scattering~\cite{Maret},  but a  short time scale  of these  methods cannot extracts information of its long time properties. This leads to the conclusion that the computer simulation of molecular dynamics  is the most suitable  tool for the study of   translation dynamics in molecular systems~\cite{Fischer,Chen}. The current understanding of  molecular dynamics in G/W mixtures is derived from  experimental results in combination with   computer simulations. Simulations can reproduce some macroscopic physical properties, but they  largely depend on the chosen models. Thus,  conclusions derived from  models regarding molecular structure and dynamics  remain uncertain~\cite{Towey}. 

A lot of effort has been devoted toward understanding of the molecular translation dynamics in water, glycerol and G/W mixtures  by measuring  the self-diffusion  coefficient, $D$.  Well known  are  the studies of molecular dynamics in liquids  by measuring $D$ in water by tracer technique~\cite{Mills},  the measurements  in the G/W mixtures by using the  interferometric micro-diffusion method~\cite{Nishijima} and by NMR methods~\cite{Tomlinson,DErrico}. The results of tracer technique are commonly used to calibrate the diffusion  measurements done by other techniques, especially those obtained  by the NMR gradient spin echo method~\cite{Hahn,Carr}. This method uses the magnetic field gradient $\nabla|\bf{B}|=\bf{G}$ (MFG) to detect the translation displacement of molecules via uneven precession of their atomic nuclear spins. A variant of this method, the pulsed gradient spin-echo (PGSE), provides   the signal attenuation proportional to the molecular mean squared displacement (MSD)  in the interval between two consecutive  MFG pulses~\cite{Torrey56,Stejskal652}. However, the PGSE measurements in water at different  pressures and  temperatures ~\cite{Krynicki,Yoshida} show values of $D$ scattered beyond the experimental uncertainty~\cite{Mills}. Differences are commonly assigned to    inaccurately calibrated MFG or to the convection flows in liquids. However, differences may also be due to the failure to observe properly some of experimental parameters as shown in the following.  In the measurement  of G/W mixture by the stimulated PGSE  technique~\cite{Mallamace}, only $D$ of  water is obtained, since its contribution to the NMR signal is well distinguished by glycerol.The study proves the  validity of the Arrhenius behavior and the Stockes-Einstein relationship between $D$ and  viscosity $\eta$ in the range of around room temperatures, but with deviations   at  temperatures close to the glass transition. The diffusion in G/W mixtures was measured also with the PGSE-FT NMR method by measuring the NMR hydrogen signal of  $CH_2$,  and  $OH$  to which both water and glycerol contribute~\cite{DErrico}. Since the proton exchange between hydroxy groups of glycerol and water is much faster than the time intervals of  spin-echo sequence, $D$ of water  component was calculated by taking into account numbers of $OH$ groups belonging to  water and glycerol molecules. The mutual diffusion coefficient of G/W mixtures was also measured by the Gouy interferometric technique, but these results return  substantially higher $D$ than obtained by  the interferometric micro-diffusion method~\cite{Nishijima} and lower than  measured by the holographic interferometry~\cite{Ternstrom}. No evident reason for these discrepancies has been found~\cite{DErrico}.

According to  the Einstein definition~\cite{Einstein}, $D$ is a time derivative of the molecular mean squared displacement (MSD) in the long time limit. When measured in a  finite time interval, as in the case of the PGSE methods, $D$ can exhibit time dependence, because  the initial velocity of  labeled molecule, ${\bf{v}}(t)$, may not be forgotten fast enough. Thus, $D$ obtained by this method may be  different from that obtained  by tracer techniques.  According to the Green-Kubo formula~\cite{Kubo2} the measurement in the finite time interval gives
\b
{D}_{zz}(\tau) = \int_0^{\tau} \ave{
{v}_z(t){v}_z(0)}_{\tau} dt=\frac{2}{\pi}\int_0^\infty D_{zz}(\omega)_{\tau}\frac{\sin(\tau\omega)}{\omega}d\omega,
\el{gk}
which is the time-dependent  self-diffusion coefficient, ${D}_{zz}(\tau)$ for the  motion along $z$-dirrection in the diffusion interval $\tau$. It is related to  the power spectrum i.e. the velocity autocorrelation spectrum  (VAS)  $$D_{zz}(\omega)_{\tau}=\int_0^{\tau} \ave{{v}_z(t){v}_z(0)}_{\tau}\cos{(\omega t)}dt,$$ where  $\langle...\rangle_{\tau}$ indicates the ensamble average over the particle trajectories in the  interval $\tau $. 

Theories ~\cite{Landau,Vladimirsky,Giterman,LandauB} and  simulations \cite{Alder1,Alder2,Visscher}  predict a long-time asymptote  of the VAF as the  $t^{-3/2}$-long time tail in liquids. Recent studies by the NMR modulated gradient spin-echo (MGSE) method~\cite{moj18} have  shown a deviation from this asymptotic properties  in simple liquids such as water, ethanol and  glycerol due to inter-molecular interactions.  In these measurements the unusual heterogeneity of molecular motion is observed, when the measurement interval is very short, which one cannot describe by a simple diffusion coefficient.  In order to enlighten these phenomena and the function of glycerol  as a colligative solute we set out to study the molecular self-diffusion in G/W mixtures by the NMR MGSE method presented in the following. 
  
\section{NMR gradient spin-echo method}

In liquids rapid molecular motion on the time scale  of pico- or nanoseconds completely nullifies the spin dipole-dipole and the first order quadrupole  interactions, while the spin interactions with electrons in molecular orbitals and the  electron mediated spin-spin interactions cannot be ignored. They appear as the chemical shifts and  $J$-couplings in the  NMR spectrum. Fluctuation of these interactions can be  characterized by  correlation functions of relevant physical quantities and affect  spin relaxation. In the gradient spin echo method,  the applied MFG is commonly strong enough that the effect of fluctuation of molecular translation velocity  prevails in echo attenuation over all other interactions.  NMR gradient spin echo sequences create the spatial spin  phase discord described by its wave vector, $\vec{q}(t)=\gamma\int_0^t{\vec{G}(t')f_\pi(t')dt'}$, in which $\pi$-RF pulses  switch $b(t')$ in  $f_\pi(t)=\int_0^t\cos{(b(t'))}dt'$  between $\pm\pi$ ~\cite{moj16}. Whenever  the molecular displacements within the interval of phase modulation are shorter than $1/\abs{\vec{q}}$, the decay of signal of the spin-echo peak at the time $\tau$ can  be approximated by the cumulant series in the Gaussian approximation as~\cite{moj81,mojcall,moj202} 
\b
 E(\tau)&=&\sum_iE_{oi}\ex{-i\alpha_i(\tau)-\beta_i(\tau)}.
\e 
Here the sum goes over the sub-ensembles of spins with   identical dynamical properties. The phase   shift 
\b
\alpha_i(\tau)=\int_0^{\tau}\vec{q}(t)\cdot\ave{{\vec v}_i(t)}dt,
\e
 can be neglected, when the averaged  velocity of spin bearing particle is zero, $\ave{{\vec v}_i(t)}=0$. This it is not in the case of  collective  molecular motion or  diffusion in non-homogeneous systems like  porous media~\cite{moj202}.

 Spin-echo attenuation is given by~\cite{mojcall}
 \b
\beta_i(\tau)&=&\frac{1}{\pi}\int_{0}^\infty \vec{q}( \omega ,\tau )\vec {D}_i (\omega ,\tau) \vec{q}^* (\omega ,\tau ) \,d\omega,
\el{att}
where  ${\vec q}(\omega,\tau)$ is the the spectrum of the spin phase discord $\vec{q}(t)$~\cite{moj18}  and where the VAS  is
\b
\vec{D}_i(\omega,\tau) = \int_0^{\infty} \ave{
\vec{v}_i(t)\otimes\vec{v}_i(0)}_\tau \cos(\omega t) dt. 
\el{psd}
According to Eq.\ref{att}, the measurement of liquid by the pulsed gradient spin echo (PGSE) sequence, which consists  of two MFG pulses of  width $\delta$ and separated for $\Delta$,   gives  the spin echo attenuation in liquids~\cite{moj13_1}
\b
\beta_i(\Delta,\delta)=\frac{ \gamma^2G^2}{\pi}\int_0^\infty D(\omega)\left(\frac{4\sin(\omega\delta/2)\sin(\omega\Delta/2)}{\omega^2}\right)^2d\omega,
\el{PGSE} 
from which  $D(\omega)$ and thus $D(t)$ can be extracted with considerable difficulty by changing $\Delta$ and $\delta$ if they are in the range of  VAF correlation time $\tau_c$. At the present state of the art, the MFG coil induction limits the  width and the shape of the MFG pulses to above  $1$ ms, which is close to or slightly above values of  $\tau_c$  in some liquids~\cite{moj18}. Thus, neglecting  the  dependence of the echo decay on $\delta$  and $\Delta$  according to Eq.\ref{PGSE}, when measuring the self- diffusion in water by PGSE method,  gives an apparent self-diffusion coefficients that may differ from one another~\cite{Mills,Krynicki,Yoshida} and also  deviate from those obtained from the theory and the  simulations of molecular dynamics  and water binary mixtures~\cite{Mahoney, Fischer00}. 

The determination of the long time asymptotic properties of VAF in dense systems is still a challenge that can be tackled by a NMR method, which directly probes the VAS. Such method is the modulated gradient spin echo~\cite{mojcall}, where  the sequence of  RF-pulses and MFG modulates the spatial dispersion  of the spin phase. The method is  basically   a Carr-Purcell-Meiboom-Gill sequence  (CPMG) consisting of initial $\pi/2$-RF-pulse and the train of $N$  $\pi$-RF pulses separated by  time intervals $T$~\cite{Carr,Meiboom}, which  was used initially  to reduce the effect of diffusion in the measurement of $T_2$ relaxation. Detailed analysis shows that the sequence   imprints  information about VAS, when applied  in the combination with MFG~\cite{moj81,moj85}. In  first applications of MGSE method,   pulsed or oscillating MFG were used to measure  water flow  through  porous material~\cite{mojcall3} and  molecular restricted self-diffusion  in porous media~\cite{moj001,Codd,Topgaard,Parsons}. It was also demonstrated how the MGSE sequence improves  the resolution of the diffusion-weighted MR images of the brain and the MRI of the  diffusion tensor of neurons~\cite{Aggarwal}.  As in the case of  PGSE, the frequency range of  MGSE  with the pulsed MFG is limited to below $1$ kHz  due to the gradient coil self-inductance. With the development of the MGSE technique in constant MFG,  the frequency induction  limit was avoided. New technique enables  higher  frequency limit, which  is now determined by the power of the RF transmitter  and the magnitude of the MFG, while the lowest limit is inversely proportional  the spin relaxation time. The use of the MGSE techniques with  fixed MFG was first  proposed in reference~\cite{mojcall} with  concern about the side effects of using RF pulses in the presence of  background MFG. Analysis of adverse interferences of both fields~\cite{moj16,moj18} shows that at suitable experimental conditions   the  MGSE signal decay  can be described as
\b
E(\tau,\omega_m)&=&\sum_i{E_{oi}\ex{-\frac{\tau}{ T_{2i}}-\frac{8\gamma^2G^2}{\pi^2 \omega_m^2}D_{zzi}(\omega_m,\tau)\tau}}. 
\el{dusapprox}
Here,  $\tau=NT$ and $D_{zzi}(\omega_m,\tau)$  denotes the component of  VAS tensor of the $i$-th spin sub-ensemble  in the direction of applied MFG at  the modulation frequency $\omega_m=\pi/T$ averaged over the interval $\tau$, and where  $ T_{2i}$ is the spin relaxation time.  The advantage of the new MGSE technique was demonstrated by  measuring the VAS  of  restricted diffusion in  pores smaller than 0.1 $\mu$m \cite{moj16}, by measuring  the VAS of  granular dynamics in fluidized granular systems \cite{Lasic06} and  by the  discovery of a new low frequency mode of motion in  polymer melts~\cite{moj14_1}. Instead of using the externally applied MFG, this MGSE method  also  allows the exploitation of  MFG generated by the  susceptibility differences on interfaces in porous systems  to obtain information about the pore morphology  and  distribution of internal MFG~\cite{moj16}. 

\section{Experiments}
NMR spectrometer with $100$  MHz proton Larmor frequency  equipped with the Maxwell gradient coils to generate MFG in steps to the maximum of $5.7$ T/m was used to measure the VAS of  in G/W mixtures at room temperatures by the MGSE method. Only water echoes are traced, as the contribution of glycerol to signal decay can be neglected due to its slow diffusion rate and short spin relaxation time. Results were checked by repeated measurements of the same samples on the   NMR-MOUSE~\cite{Bluemich} operating at $18,7$MHz proton Larmor frequency with  fixed MFG of $21.6$ T/m. The high magnetic  field of the $100$ MHz spectrometer allows  measurements with a high  signal to noise ratio, but with the top frequency  limited  to $3$ kHz due to the weak MFG. On the other hand, the large MFG of one-sided magnet of the NMR MOUSE allows  measurements of very slow diffusion  in the   frequency range up to about $10$ kHz. However, its fix MFG limits  the measurements on G/W mixtures to below 1 kHz  due to fast diffusion rate of water.

Samples  of pure glycerol  (99.5$\%$-Sigma-Aldrich),  distillate water and   mixtures with several different  volume fractions of glycerol were prepared in plastic bottles  of  $100$ ml volume. Tightly sealed with a plastic lid were kept  for several days before  used for the measurements. The samples  were loaded into $15$ mm long and  $5$ mm wide pyrex glass ampules and closed with paraffin tape to be inserted in the head of the $100$  MHz NMR spectrometer.  While the flat shelf of one-sided magnet of  the NMR-MOUSE allows the measurements of mixtures in bottles without the loading step.  In order to avoid a possible impact of restricted diffusion, containers with the diameter much larger than the molecular displacements  in  the time of  $100$ ms long  intervals of  measurement were used.  In addition, the effect of restricted diffusion is even further reduced by the initial $\pi/2$-RF pulse of the MGSE sequence applied in the background of MFG, because it excites only a few mm narrow slice of sample. 
 \begin{figure}
\centering \scalebox{1}{\includegraphics{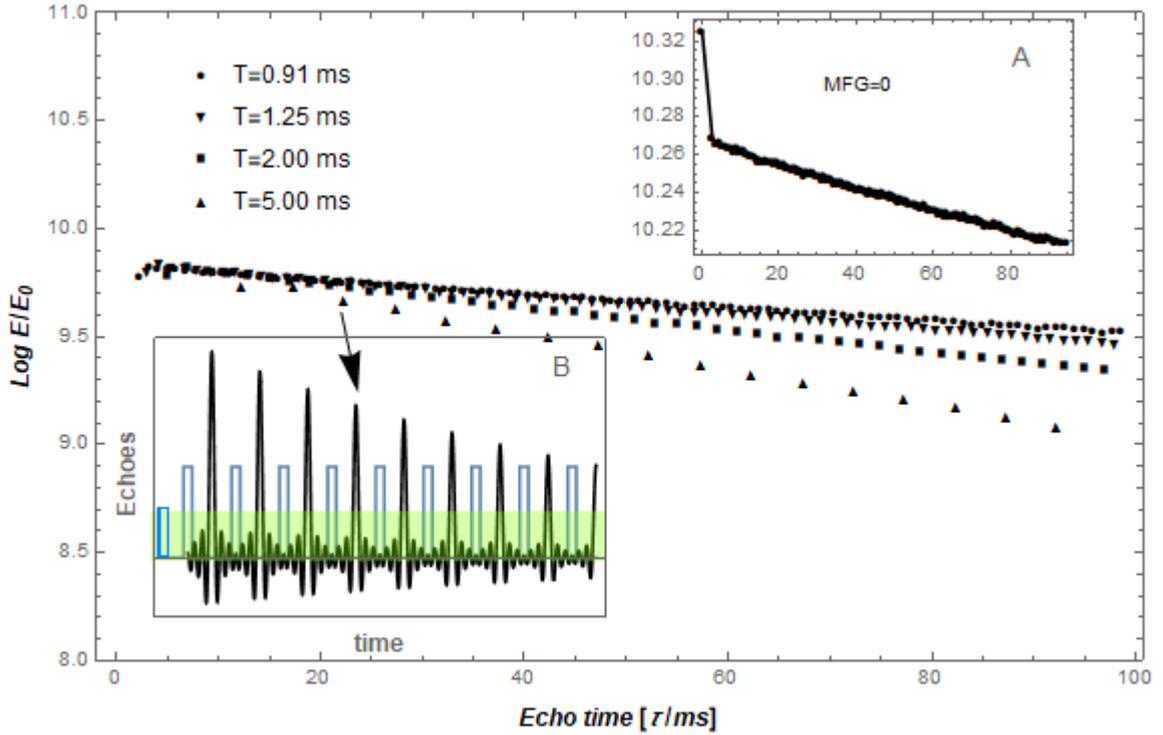}}
\caption{MGSE spin echo decays in a mixture of $20$ vol$\%$ glycerol in water  at  different modulation frequencies ($\omega=\pi/T$). The inserted images show: A) the signal decay in zero MFG and B) MGSE sequence with RF pulses (blue) and MFG (green) giving spin echoes (black) whose peaks contain information about molecular motion.  \label{fig1}}
\end{figure}
 \begin{figure}
\centering \scalebox{0.65}{\includegraphics{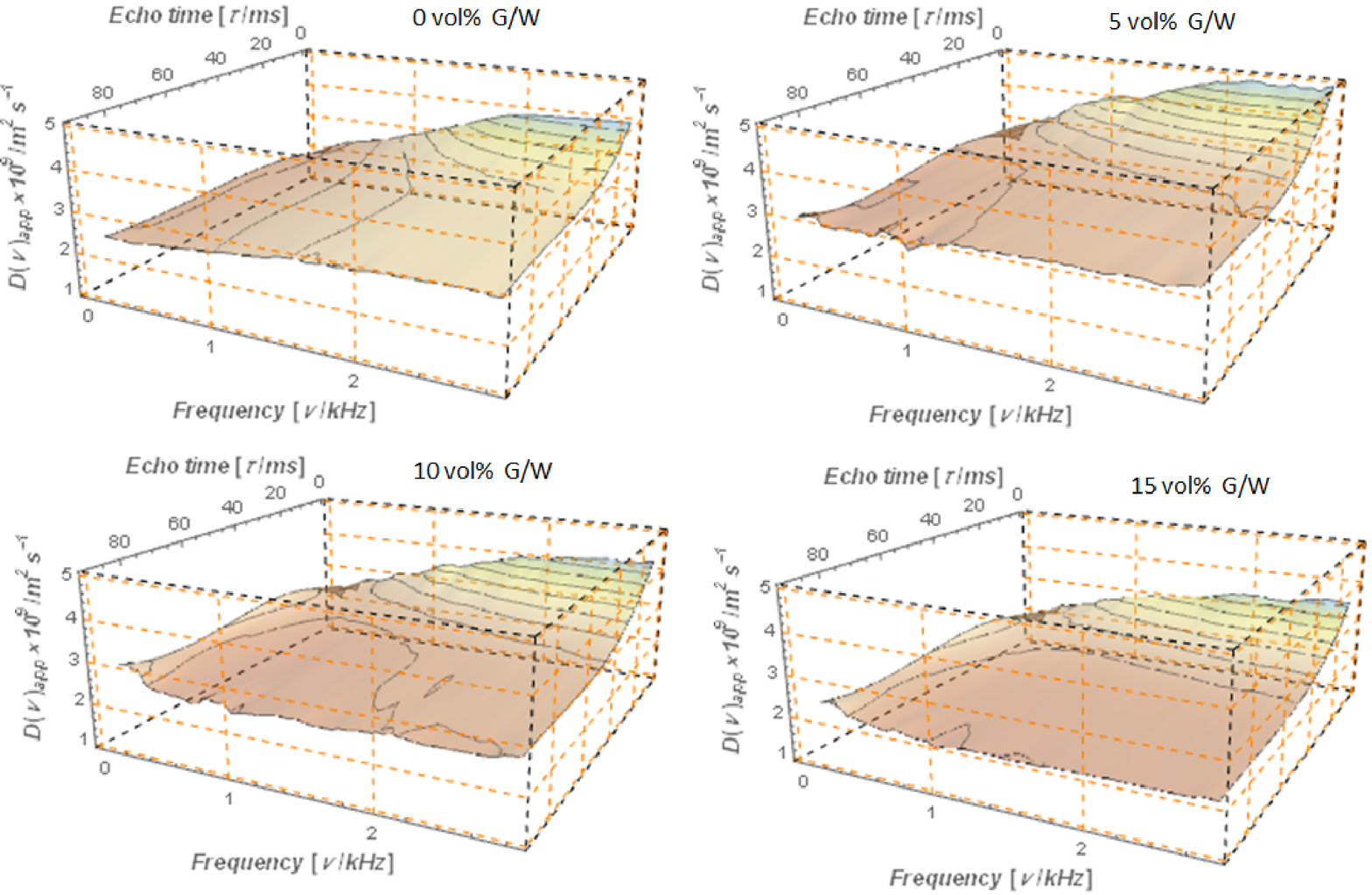}}
\caption{3D temporal-frequency plots of  apparent VAS for the G/W mixtures with $0$ vol$\%$, $5$ vol$\%$, $10$ vol$\%$, $15$ vol$\%$ of glycerol. \label{fig2}}
\end{figure}
 \begin{figure}
\centering \scalebox{0.8}{\includegraphics{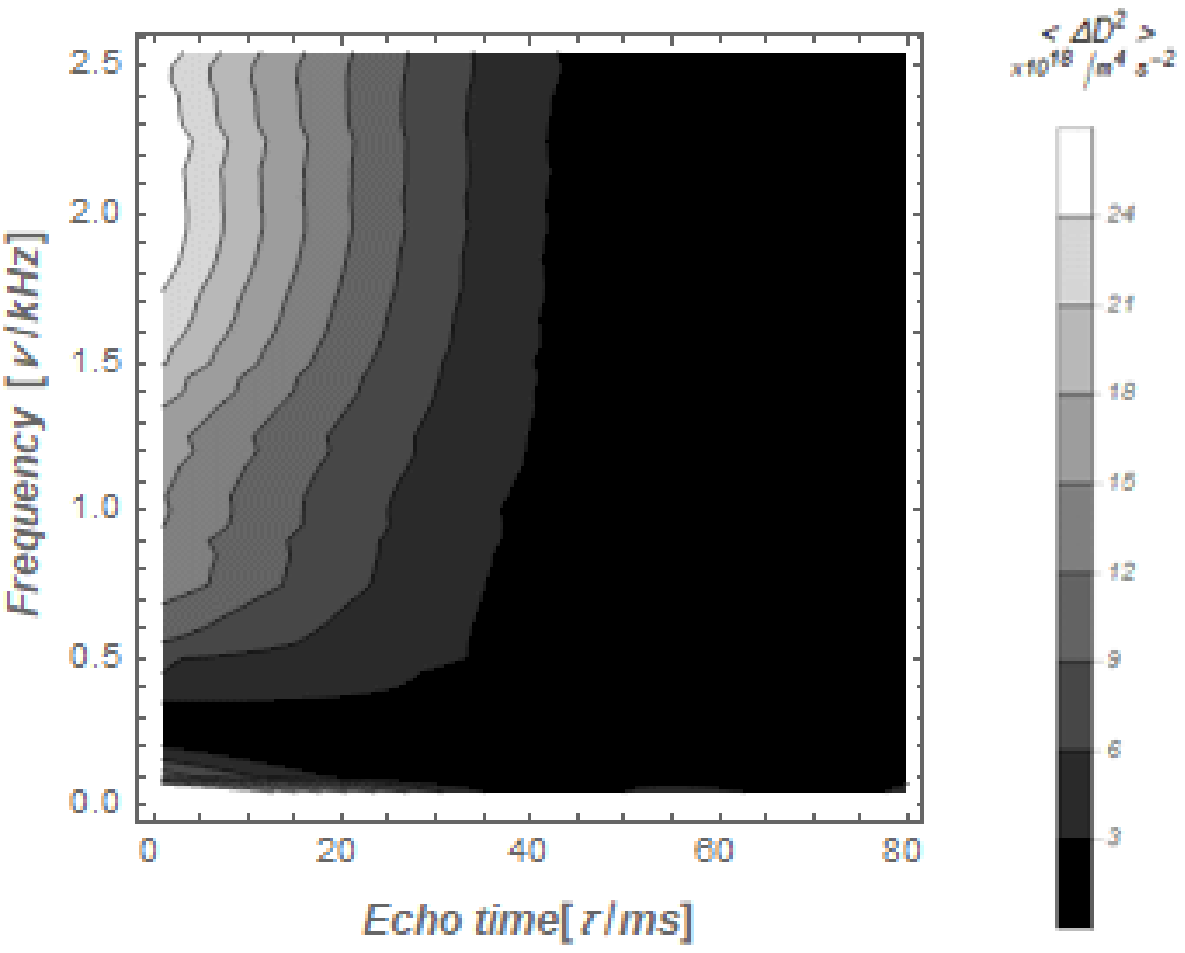}}
\caption{The contour plot  of the distribution variance  of the diffusion coefficient, $\ave{\Delta D^2}$,  for the G/W mixture with $5$ vol$\%$ of glycerol showing the diversity of molecular motion observed only at short observation times and at high modulation frequencies.\label{fig3}}
\end{figure}

In the measurements on the $100$ MHz NMR device,  the amplitudes of echo peaks  are recorded in the time interval of $100$ ms. Measurements are repeated with different  $T$ and  magnitude of MFG, $G$, in a way to keep the product $T\times G=const$.  Fig.\ref{fig1} shows that the  number of recorded peaks increases with  the shortening of $T$. In this way  the exponential decay of $N$-echo peaks is effected only by the changes of  $D(\omega,\tau)$ at the frequency $\omega=\pi/T$, where $\tau$ denotes average over the  interval  $\tau=NT$ . The contribution of spin relaxation to the signal decay is determined  by  separate measurements in zero MFG as  shown in the inset picture in the Fig.\ref{fig1}. The spin relaxation  exhibits almost a clear mono-exponential decay, except in the initial interval of a few milliseconds, with the  rapid decay believed  to belong to  water bounded in clusters..

\section{Results and discussion}

Data was analyzed in the following way. A fifth order polynomial is fitted to the series of echo peak amplitudes with given $T$  as a function of time $\tau$. The time derivative of this fit represents the diffusion spectrum if the the the spin relaxation is accounted for by proper normalization of the echo decay. The MGSE measurements of G/W mixtures on the $100$ MHz NMR device give   time dependencies of spin-echoes $E(\tau)$ that deviate from the anticipated  mono-exponential  decays particularly  in pure water and G/W mixtures with  lower glycerol contents. The deviations are clearly visible in the 3D frequency/temporal  plots  $D(\omega,\tau)$ of  the time derivatives of curves obtained by fitting the echo peaks with the coefficient of determination of $R^2>0.99999$  shown in  Fig.\ref{fig2} and Fig.\ref{fig6}. Instead of  constant value  along the echo time, $\tau$, anticipated for the mono-exponential decays, discrepancies occur that cause the 3D plots to show a curved surface at short intervals $\tau$ and of higher frequencies. In the case of simple liquids~\cite{moj18}, we interpreted similar curved surface  by the molecular diffusion diversity due to the motion in the vortexes of hydrodynamic fluctuations. Namely, during initial  interval after the spin excitation, when the trajectories of the molecules are still short, the spins may observe local inhomogeneity caused by the distribution of internal MFG in the porous medium~\cite{moj18} or motional diversity due to fluxes or hydrodynamic oscillation in liquids~\cite{moj16}. By grouping the spin bearing particles  into separate sub-ensembles corresponding to spins with the different spin-echo attenuation, we can describe  the  induction signal  with the distribution function $P(D)$ as $E(\tau)=\int P(D) e^{-s D\tau} dD$ with  $s= \frac{8\gamma^2G^2}{\pi^2\omega^2}$  given in Eq.\ref{dusapprox}. In the case of narrow distribution, the signal attenuation can be   approximated by
\b
\log E=\beta(\tau)\approx -\tau/T_2-s\ave{D}\tau+\frac{s^2}{2}\ave{\Delta D^2}\tau^2+... 
\e
 Here $\ave{ D}$ is the mean  diffusion coefficient and $\ave{\Delta D^2}$ is the variance of distribution. In the cases of a non-exponential decay the time derivative of echo attenuation does not give VAS, but some apparent one, $D_{app}(\omega,\tau)$, which conveys  information on the mean diffusion coefficient and its distribution.  3D frequency/temporal  plots of $D_{app}(\omega,\tau)$   for pure water and for G/W mixtures with $ 0.05, 0.10$ and $0.15$ volume fraction of glycerol are shown in Fig.\ref{fig2} and  for the G/W mixtures with $0.20, 0.33, 0.50$ and $0.66$ volume fraction of glycerol in  Fig.\ref{fig5}.  In  water and  mixtures with the lower glycerol content a curved surface is evident in the 3D plots of spectra in the range of  short $\tau$ and at higher frequencies due to  the distribution of  the self-diffusion coefficients.  The part of curved surface is clearly visible in the contour plot of the  second derivative of $\beta(\tau)$ of the mixture with $5$ vol$\%$ of glycerol content in Fig.\ref{fig3}, which  shows how $\ave{\Delta D^2}$ occurs at high modulation frequencies and disappears  with increasing echo-time at $\tau> 40$ ms, when the  trajectories of spins become long enough to span the whole extend of heterogeneity and to average off the diffusion diversity into $\ave{\Delta D^2}=0$,  what is also seen in Figs.\ref{fig2} and \ref{fig6}. Similar diversity was observed in the MGSE measurements  of VAS in water,  ethanol and toluene with the explanation of the  molecular self-diffusion  in the vortexes of  hydrodynamic fluctuations~\cite{moj18}. 

By increasing the glycerol content, the curved surface of spectra is reduced, but in the mixtures with $\geq10$ vol$\%$ of glycerol  a new  spectral feature appears  in the form of  a low frequency ridge that levels at higher frequencies.   In order to enhance visibility of spectral changes and to avoid the part affected by hydrodynamic fluctuation, where $\ave{\Delta D^2}\not=0$,  we present  in Fig.\ref{fig4} the VAS   at echo  times  $\tau>50$ ms for the  mixtures with $0.00, 0.05, 0.10$ and $0.15$ vol$\%$ of  glycerol. The spectrum  of mixture with $5$ vol$\%$ of glycerol has a form  similar to that of pure water, only shifted up-wards by about $ 30 \%$. The upwards shift indicates a disruption of  hydrogen bonding in water caused by  a low glycerol content in water.   A drastic change of spectra appears at the concentrations of $10$ and $15$ vol$\%$ of glycerol indicating  new type of inter-molecular interaction, which has a strong impact on the molecular dynamics. 3D presentations of spectra in Figs.\ref{fig2} and \ref{fig5} clearly show how the  increase of glycerol content lowers the spectrum level, reduces the spectral hump, i.e. the curved surface at short $\tau $, which is attributed to the bulk water. The low frequency  ridge of the new spectrum, which starts to appear at $10$vol$\%$ of glycerol, remains almost unchanged with the increase of glycerol content.

In order to ensure the validity of results obtained by the high frequency NMR device, the measurements of the samples were repeated by using the  $18.7$ MHz NMR MOUSE device. Its strong and fixed MFG enables the measurement of samples up to about $10$ kHz,  but with the lower S/N ratio due to lower Larmor frequency. It also does not permits the   measurements of water and G/W mixtures below $1$ kHz, due to excessive spin echo attenuation.  Fig.\ref{fig6} shows the  results of measurements  on both devices, which match  well in the overlapping  frequency range.
\subsection{Water-cluster interaction} 
In references\cite{Gadre,Guo,Vaitheeswaran,Saykally} a formation of clusters in the G/W mixtures was explained by the energies of  bonding between water molecules and between glycerol molecules, which are lower than  between water and glycerol molecules. Thus, the new spectral features of G/W mixture can be attributed to the formation of clusters  around  the hydrofilic glycerol molecules, and this interaction with unbound water changes molecular dynamics.  The  disappearance of  spectral hump with the increase of glycerol content indicates the depletion of the  free water basin through the formation of new clusters. We believe that there are two contributions to the mixture VAS, when measuring proton signal by MGSE method: The contribution of free water that is  not in contact with the clusters, and whose content decreases with the increase of glycerol concentration, and  the  water that  interacts with the clusters or even could  exchange with the clustered water. The VAS of pure glycerol and  water bounded in the clusters  cannot be detected due to a slow diffusion rate~\cite{moj18} and the short relaxation time, and is only partially observed in Fig. \ref{fig1} for the spin-echo decays in zero MFG. 

At room temperatures the  self-diffusion coefficient of G/W mixtures  does not deviate significantly  from Arrhenius's law~\cite{DouglassVFC}, allowing  to consider the  molecular diffusion  as the motion of particles  occasionally caught up in potential wells created by their neighbors. Quite commonly the molecular dynamics is described  by  the Langevin equations, where the effect of the environment is taken into account through the fluctuating and frictional forces. The method has been widely used in the study of structural and thermal properties of matters in different phases~\cite{Snook}. However,  the  inter-molecular interaction and the coupling to other degrees of freedom are difficult to effectively  include into the description. The description of   interaction between  water-water, water-glycerol and glycerol-glycerol molecules  with  a simple Lennard-Jones potential turned out to be unsuccessful even though the potential is modified  by the bifurcation of the single minimum into two or more minima~\cite{Stanley}. However, the results of MGSE measurements in ordinary fluids ~\cite{moj18} showed that the results can be successfully explained by Langevin equations (LE), in which we consider that molecular motion at high temperatures alters molecular interactions to such an extent that they can be approximated by harmonic interaction ~\cite{Vinales}. This means that in mutual collisions, the molecules occasionally get caught in a harmonic well, which can be described by a constant $k$ and with a minimum at $a(t)$ between $x_i$ and $x_j$ locations of colliding molecules. This allows to describe the molecular dynamics by the  set of coupled LE 
\b
m_i\frac{d^2x_i}{dt^2}+\gamma\frac{dx_i}{dt}+k\sum_{j\not=i}^n{(x_i-x_j-a(t) )}=f_i(t)
\el{lang}
in  which  the $i$-th particle is coupled to $n$ of its  closest neighbors. Here, $m_i$ is the particle mass, $\gamma$ is the friction coefficient, and $f_i$ is the random force. The friction and the random force represent two consequences of the same physical phenomenon and are interrelated ~\cite{Kubo2}
\b
\gamma(\omega)=\frac{1}{k_BT}\int_0^\infty{\ave{f(t)f(0)}\exp{(i \omega t)}dt},
\e
where $k_B$ is the Boltzmann constant. Neglecting  the fluctuation of $a(t)$ and the inertial terms of LE at low frequencies , when $\omega<k/m_i$, and assuming that all particles are subjected to the same friction  $\gamma(\omega)\approx\gamma=\frac{1}{k_BT}\ave{|f|^2}$,  the solution gives the VAS of harmonically coupled water molecules, $D_{w}(\omega)$,  in the form 
 \begin{figure}
\centering \scalebox{0.6}{\includegraphics{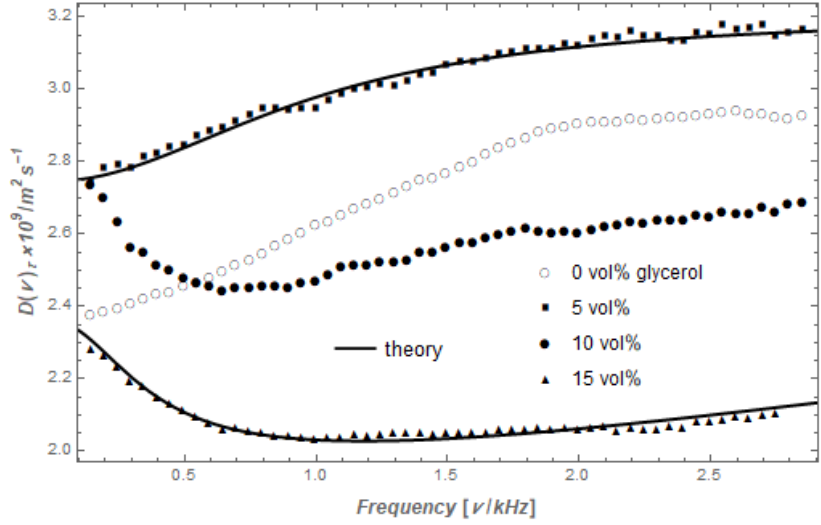}}
\caption{The dots show  VAS  of G/W mixtures measured at echo  times  $\tau>50$ ms and  the fitting  curves  correspond to  the harmonic couplings of diffusing particles described with  Eq.\ref{coup} for  $5$ vol$\%$ and with  Eq.\ref{lanWG} for  $15$ vol$\%$  of glycerol content . \label{fig4}}
\end{figure}
\b
 D_{w}(\omega)=\frac{k_BT}{\gamma}\frac{n+\tau_c^2\omega^ 2}{n^2+\tau_c^2\omega^ 2}.
 \el{coup}
 where the correlation time is $\tau_c=\gamma/k$.  At zero  frequency,  the Einstein diffusion coefficient, $D_{w}(0)=\frac{k_BT}{n\gamma}$, depends on the number of coupled molecules, while at high frequencies, $D_{w}(\infty)=\frac{k_BT}{\gamma}$ is the diffusion rate  of molecules  escaping the inter-molecular capturing. This formula provides a good  fit to the VAS of pure water, glycerol and  ethanol~\cite{moj18} as  well  also to the results of our measurements for the VAS of  G/W mixture with $0.05$ vol$\%$  glycerol shown  in  Fig.\ref{fig4}.  However,   $D_{w}(\omega)$  cannot fit  to the VAS of G/W mixture with  the glycerol concentrations equal to or greater than $10$ vol$\%$  as shown in  the same figure. In the latter case, one must consider  the interactions of water with the water clustered around glycerol. With a slight exaggeration that water molecules and clusters experience the same friction, $\gamma$, that the coupling constant between small and large particles is the same $k$, and given that the mass of clusters is large enough that  the  inertial term in their Langevin equations cannot be neglected at low frequencies, i.e.  $\omega\approx\gamma/M$, the solution for the system of  $n_w$ light water molecules and $n_c$ heavy clusters gives   the VAS  for the  water molecules interacting with clusters as
\b
D_{wc}(\omega)=\frac{k_BT}{\gamma  n_w}\left(\frac{\left(n_w-1\right)\tau ^2 \omega ^2 }{
   \left(n_c+n_w\right)^2+\tau ^2 \omega ^2 }+\frac{n_c n_w+\tau ^2 \omega ^2+\left(n_w-\frac{\omega^2}{\omega_o^2}\right)^2}{\left(2 n_c+ n_w\right)n_c +\tau ^2 \omega ^2+\frac{ n_c^2 \omega ^2}{\tau^2\omega_o^4}+\left(n_w-\frac{ \omega ^2}{ \omega_o^2}\right)^2}\right),
\el{lanWG}
and the VAS for the clusters interacting with water as
\b
D_{cw}(\omega)=\frac{k_BT}{\gamma  n_c}\left(\frac{\left(n_c-1\right)\tau ^2 \omega ^2 }{   \left(\frac{ \omega ^2}{ \omega_o^2}-n_c-n_w\right)^2+\tau ^2 \omega ^2 }+\frac{\left(n_c +n_w\right)n_w+\tau ^2 \omega ^2}{\left(2 n_c+ n_w\right)n_c +\tau ^2 \omega ^2+\frac{ n_c^2 \omega ^2}{\tau^2\omega_o^4}+\left(n_w-\frac{ \omega ^2}{ \omega_o^2}\right)^2}\right),
\el{lanWGC}
where $\omega_o=\sqrt{k/M}$. This may be an excessive simplification, but allows the calculation of  results that serve at least qualitatively to compare with our experimental results.  Fig.\ref{fig4} shows a good  fit of $D_{wc}(\omega)$  to the data for the VAS of mixtures with  $15$ vol$\%$ of glycerol, if assuming that each water molecule interacts with one of adjacent aqueous molecules and one cluster  and that the interaction between clusters is neglected .

\subsection{Rate thickening of molecular friction  in G/W mixture}

 \begin{figure}
\centering \scalebox{0.65}{\includegraphics{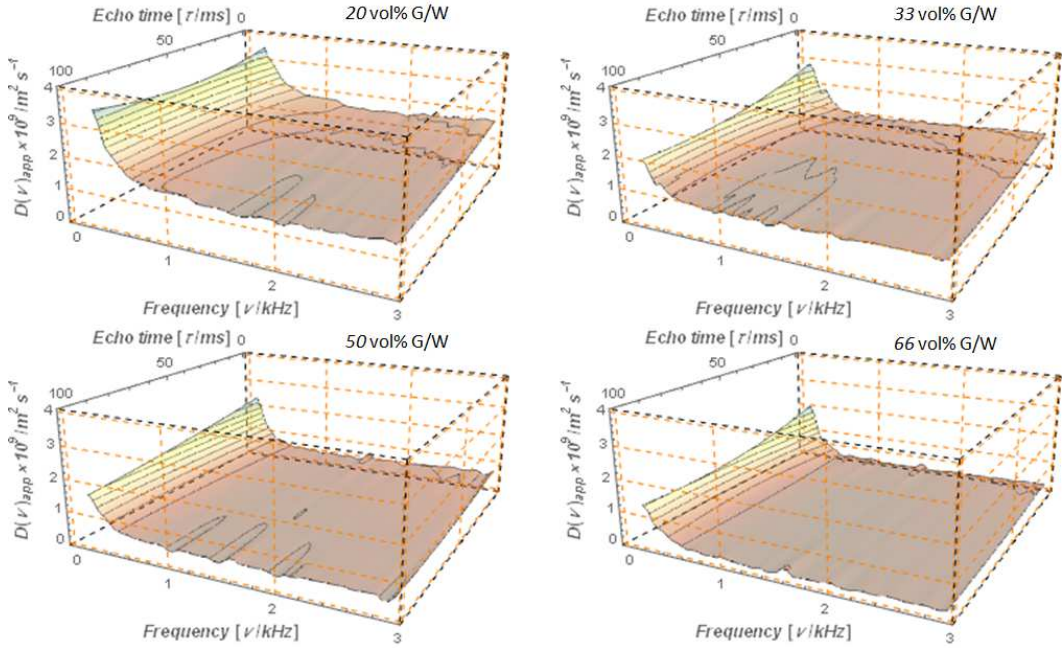}}
\caption{3D temporal-frequency plots of apparent  VAS for the G/W mixtures with $20$ vol$\%$, $33$ vol$\%$, $50$ vol$\%$, $66$ vol$\%$of glycerol\label{fig5}}
\end{figure}
 \begin{figure}
\centering \scalebox{0.9}{\includegraphics{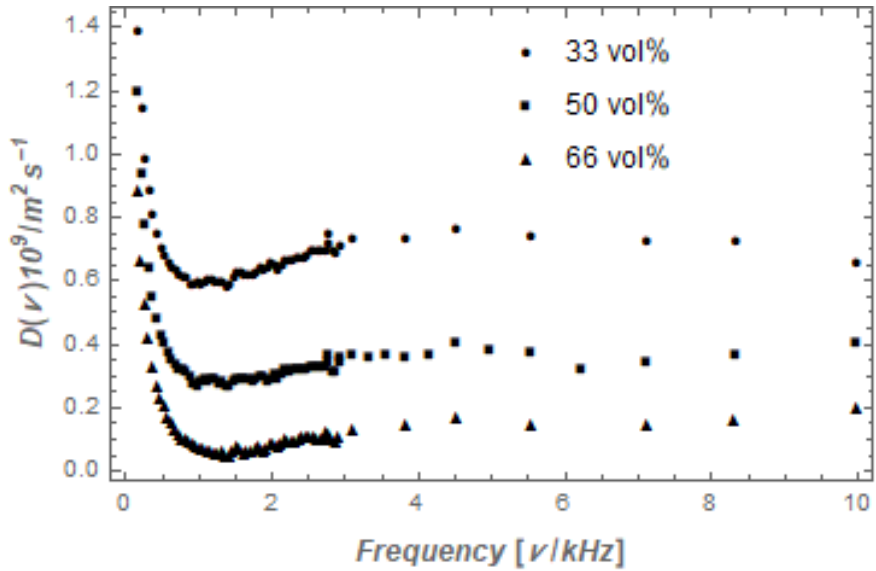}}
\caption{ VAS of G/W mixtures obtained  by MGSE method on  the $100$ MHz NMR set-up (frequencies below $3$ kHz) and on the $18.7$ MHz NMR MOUSE (frequencies above $3$ kHz) match well in the overlapping region. \label{fig6}}
\end{figure}

\begin{figure}
\centering \scalebox{1.3}{\includegraphics{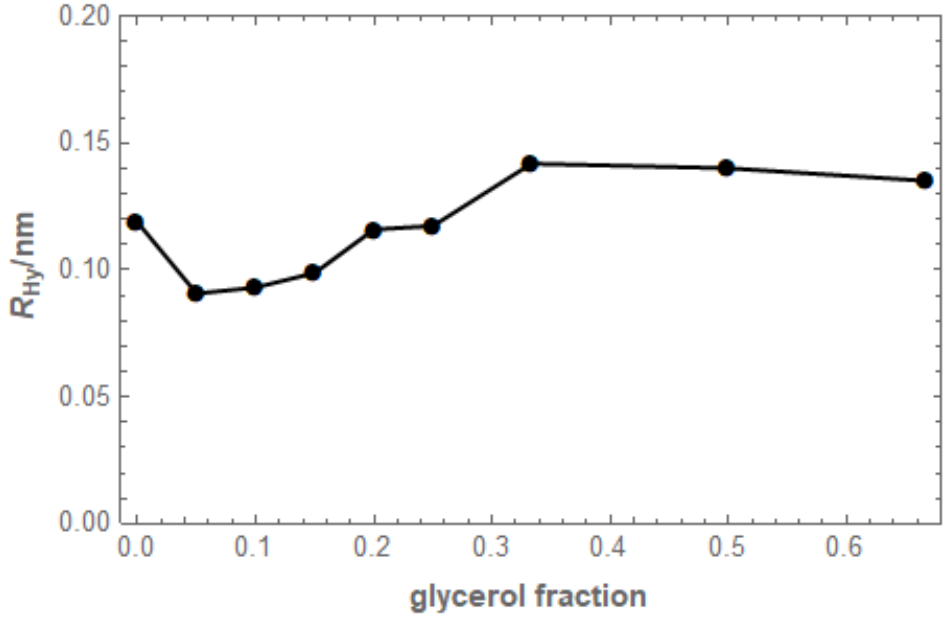}}
\caption{The size of the hydrodynamic radius of the diffusing molecules, determined according to the SE formula, proves that only the properties of  water molecules in the G/W mixture were observed.\label{fig7}}
\end{figure}

\begin{figure}
\centering \scalebox{1.3}{\includegraphics{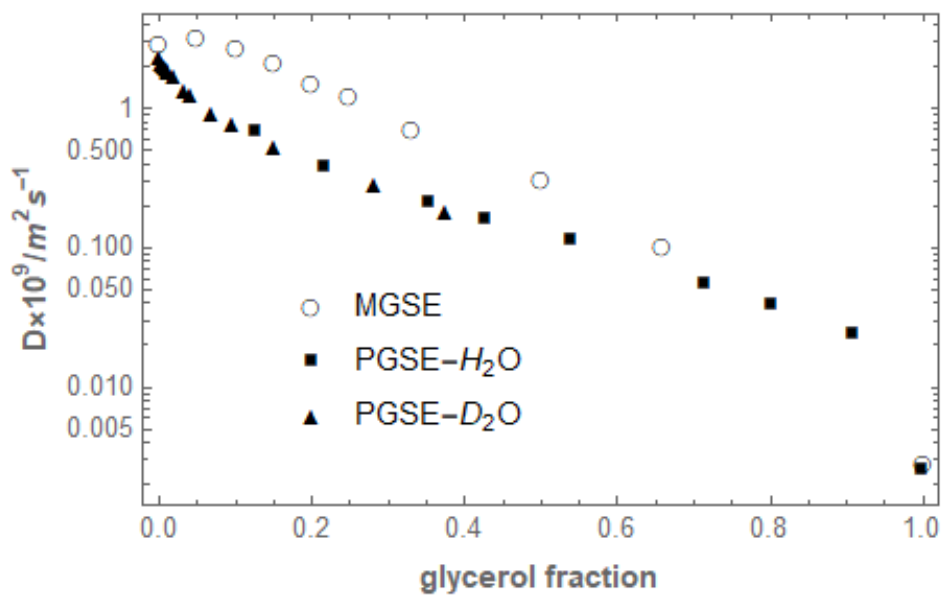}}
\caption{The diffusion coefficient of G/W mixtures measured by our MGSE method, presented together with the results of PGSE measurements from the reference~\cite{DErrico}.\label{fig8}}
\end{figure}

As shown in references~\cite{Cheng,Mallamace2} the self diffusion coefficient  of G/W mixture decreases and  the viscosity increases with decreasing temperature according the Stokes-Einstein formula (SE) in  a wider range around room temperature. This formula is derived  from the  Einstein-Smoluchowski (ES) relation~\cite{Einstein1,Smoluchowski}
\b
D=\frac{k_B T}{\gamma}
\el{ES} 
by relating the fluid  viscosity $\eta$ to the friction as  $\gamma=s\pi R \eta$, which is derived from the Navier-Stokes equations for the  spherical objects of radius  $R$ moving with  small Reynolds numbers in a  fluid. Here, the factor  $s$  depends on the boundary condition and is equal to 4 for the slip boundary condition and 6 for the stick boundaries. Thus, the hydrodynamic radius of a molecule $R$ may be derived directly from the diffusion coefficient using the SE relation, if  the viscosity of solution is known.  Fig.\ref{fig7} shows  the  dependence of water hydrodynamic radius on the glycerol content obtained by using  the diffusion coefficients  derived from the MGSE measurement, shown in Fig.\ref{fig8}, and the viscosity of G/W mixture from the reference~\cite{Chang} with the assumption of  slip boundaries. The values are close to the commonly accepted radius of water, $0.14$ nm~\cite{Yoshida} in the range of higher glycerol content, but slightly lower at low glycerol content, which can be attributed to the effect of hydrodynamic fluctuations.

In the generalized LE, hydrodynamic interactions  and the correlation between random forces at different time are taken into account by introducing   friction forces with  a memory kernel, meaning that the friction acting on the particle depends on the velocity at an earlier time  $\gamma(t)$ ~\cite{Kubo2} giving  the  VAS in the form
\b
D(\omega)=Re\frac{k_BT}{i\omega m+\gamma(\omega)}.
\e 
 which can be considered as a  generalized ES equation, in the range of low frequencies, $\omega<\gamma(\omega)/m$, or a  generalized SE relation, if  friction is expressed by a shear rate viscosity $\eta(\omega)$, $\gamma(\omega)=s\pi R\eta(\omega)$.  Generalized SE relation was used to obtain the viscoelastic module of a complex fluid from the microscopic motion of small particles~\cite{Mason} in order  to understanding the swimming of microorganisms, and the sedimentation in fluids. While the generalized SE gives good estimates for the motion of larger objects, its use for the molecular  diffusion could lead to a systematic failure. Such models tend to severely underestimate molecular radius $R$  from  the diffusion coefficients or vice versa~\cite{Evans}. 
 Since  the hydrodynamic radius  of small molecules  is  weakly dependent on  the flexibility and density of the molecular structure, we still can assume   the  proportionality between the frequency dependent molecular friction and the shear rate viscosity in simple liquids. Thus, the inverse proportionality  between  the shear rate viscosity and the VAS,   $\eta(\omega)\approx D(\omega)^{-1}$,  reveals the  shear rate thickening of viscosity of our mixtures, which appears at the glycerol contents equal to or larger than $10$ vol$\%$.  

Our interpretation with the simplified LE confirms that the thickening  is a consequence of water interaction with  the hydro-cluster formed around glycerol molecules. It corresponds to the common  understanding that the shear thickening viscosity  is related to the presence of ''clumps'' in liquids ~\cite{Kretser}.

\section{Conclusion}
In  Fig.\ref{fig8} the diffusion coefficient obtained by the PGSE method as reported in article~\cite{DErrico} and obtained here by  MGSE method are compared. In MGSE method the   VAS at the  decay time $\tau=70$ ms and  at the modulation frequency $\nu=3$ kHz  is taken as a self-diffusion coefficient  in order to avoid  the part of spectrum with the humped surface  belonging to the diffusion diversity created by the hydrodynamic fluctuations and also to be outside the low frequency spectral ridge, since its  value  cannot be exactly  determined  due to the low-frequency limit of the method.  The results of both measurements match only in the cases of pure water and pure glycerol. The differences  may be due to the different  interpretation of  spin-echo decay,  given in Eq.\ref{PGSE} for the PGSE and Eq.\ref{dusapprox} for the MGSE method. The article~\cite{DErrico} does not mention the width of the MFG pulses in their experiments. This value is crucial for the correct analysis of the PGSE measurements. In addition, the differences may also appear because the proportions of glycerol, bounded and unbounded water in the hydroxyl NMR spectral line were not properly estimated in determination of water $D$~\cite{DErrico}. Unlike in the PGSE method, in the processing of the MGSE spin-echo decay no calibration is required to match  results of other methods, but one only needs  to know the exact value of the MFG according to Eq.\ref{dusapprox}.

Study of molecular dynamics in G/W mixture by the  NMR MGSE method  unveils  the low-frequency feature of water  VAS with two indications: The small amounts of glycerol in water  only partially weakens the hydrogen bonding  network in water and  thus increases the diffusion rate, and that  the glycerol content  equal or higher than $10$ vol$\%$  brings about a new feature of VAS, which is the consequence of water interaction with  the hydro-clusters formed around hydrophilic glycerol molecules. These interactions strongly influence  the translational molecular  dynamics in liquid resulting in the  shear rate thickening of viscosity of water  in mixtures. The  shear rate thickening of viscosity  alters the dynamics of other molecules if immersed in such a liquid. Instead of using the SE relation, the effect of a liquid on a submerged molecule one can treat it more correctly  with the generalized EC equation, in which the friction spectrum of friction embodies the  interaction of a molecule in a fluid. With this in mind, instead of the shear rate viscosity thickening, we can talk about  the  rate thickening of the friction coefficient in liquids, which causes an environment where rapid molecular motions and collisions are strongly inhibited.

It is well known that any protein exists  as an unfolded polypeptide or random coil when translated from a sequence of mRNA to a linear chain of amino acids. Protein folding is the physical process by which a protein chain acquires its native 3-dimensional structure, a conformation that is biologically functional. A large number of experimental and simulation studies  tested whether folding reactions are diffusion-controlled, whether the solvent is the source of the reaction friction, and whether the friction-dependence of folding rates generally can provide insight into folding dynamics~\cite{Jacob,Hagen2}, but these  some simple questions still remain unanswered. Simulations and theory provide some insight into possible physical mechanisms of internal friction, but there are no experimental demonstrations of these ideas. An answer to this dilemma could be that in  suitable solvents such as the G/W mixtures,  the rate thickening of the friction coefficient dampens  rapid molecular motion  and collisions between molecules to create a condition  for the slow process of spontaneously  folding  of disordered poly-peptides into  biologically active protein molecules when  immersed in them~\cite{Gregory,Wang}.  

As in the cases of MGSE measurements of pure liquids~\cite{moj18}, the  VAS of  G/W mixtures with the low glycerol content  show a similarity of the diffusion  diversity  explained by the molecular self-diffusion in the vortexes of hydrodynamic fluctuation,  which disappears at higher glycerol concentrations. It can be explained that  stronger molecular  interactions in G/W mixtures cause  faster fluctuation of hydrodynamic  vortexes at a  rate that cannot be observed  at the shortest  $T$ intervals achieved by our device\cite{moj18}. 


\end{document}